\documentclass[usenatbib]{mn2e}

\usepackage{epsfig}

\begin{document}

\title{Probing quintessence: reconstruction and parameter estimation
from supernovae}

\author [Brian F. Gerke and G. Efstathiou]{Brian F. Gerke$^{1,2}$\thanks{
\texttt{bgerke@socrates.berkeley.edu}} 
and G. Efstathiou$^1$\thanks{\texttt{gpe@ast.cam.ac.uk}}\\
$^1$Institute of Astronomy, Madingley Rd., Cambridge CB3 0HA, UK\\
$^2$Department of Physics, University of California--Berkeley, Berkeley, CA 94720, USA}
\date{\today}

\maketitle
\begin{abstract}
We explore the prospects for using future supernova observations to
probe the dark energy.  We focus on quintessence, an evolving scalar
field that has been suggested as a candidate for the dark energy.
After simulating the observations that would be expected from the
proposed SuperNova / Acceleration Probe satellite (SNAP), we
investigate two methods for extracting information about quintessence
from such data.  First, by expanding the quintessence equation of
state as $w_Q(z) = w_Q(0)-\alpha\ln(1+z)$, to fit the data, it is
possible to reconstruct the quintessence potential for a wide range of
smoothly varying potentials.  Second, it will be possible, to test the
basic properties of the dark energy by constraining the parameters
$\Omega_Q$, $w_Q$ and $\alpha$. We show that it may be possible, for
example, to distinguish between quintessence and the cosmological
constant in this way.  Further, when supernova data are combined with
other planned cosmological observations, the precision of
reconstructions and parameter constraints is significantly improved,
allowing a wider range of dark energy models to be distinguished.

\end{abstract}
\begin{keywords}
supernovae: general -- large-scale structure of Universe -- 
cosmology: miscellaneous. 
\end{keywords}


\section{Introduction}

There is now strong evidence, from observations of type Ia supernovae
\citep{SCP, Hi-z} and CMB anisotropy \citep{Bernardis01,boom,dasi},
that we live in a universe that is geometrically flat and
dominated by a nearly homogeneous component with negative pressure---the 
dark energy---which is causing the cosmic expansion to accelerate.  The
existence of dark energy has recently been confirmed, independently of the
supernova data, by combining the latest galaxy clustering data with CMB 
measurements \citep*{GPEetal02, WTZ01}.  The most obvious dark energy 
candidate is vacuum
energy, represented by the cosmological constant $\Lambda$, which has
pressure $p_\Lambda = -\rho_\Lambda$.  Other dark energy candidates,
include a network of topological defects and the much-discussed possibility 
of quintessence \citep*{CDS98}, a spatially inhomogeneous, evolving component 
which is usually represented by a scalar field evolving in a potential (an 
idea first introduced by \citet{PR88}).  In this paper we will focus on
quintessence, though some of the results apply to other forms of
dark energy  in so far as these can be parameterised by a simple equation of
state.

It is our goal to explore what could be learned about the dark 
energy from future 
observations of type Ia supernovae (SNIa), 
such as may be possible with the proposed SuperNova / Acceleration Probe 
(SNAP) satellite \citep{SNAP}.  This satellite aims to observe roughly 2000 
supernovae 
a year for three years, with very precise magnitude measurements and 
negligible systematic errors, out to a redshift of $z=1.7$.  
Observations of this sort will permit a very precise measurement of
the magnitude-redshift relation $m(z)$ and hence of the distance-redshift
relation $r(z)$, which will probe the expansion history of the universe.

Current supernova observations already put weak constraints
on the dark energy density $\Omega_Q$ and equation of state $w_Q$. If the universe
is assumed to be flat (\emph{i.e.}, $\Omega_k \equiv 1-\Omega_M-\Omega_Q = 0$) and $w_Q$ is
assumed to be constant in time, we have $\Omega_Q\ga 0.5$ and $w_Q\la -0.4$
\citep{SCP}.  These constraints improve considerably when they are
combined with independent constraints from the CMB \citep{GPE99}
or large-scale structure \citep*{PTW99}---most importantly, such combined
data sets give $w_Q\la -0.6$.

Several studies have examined the improved parameter constraints
that may be possible with SNAP.  For example, \cite{WA01,WA01-2} have
found that SNAP can constrain a constant $w_Q$ to better than $10\%$ accuracy,
allowing some models to be distinguished from a cosmological
constant ($w_\Lambda=-1$).  In addition, many authors have examined
the possibility of using SNAP to distinguish the evolution of $w_Q$
with redshift \citep*{Astier00,BM00,Goliath01,HT00,MBS01,WA01,WA01-2}.  
There is a general consensus that SNAP data alone will not be able
to distinguish an evolving equation of state.  Our results are in
broad agreement with these studies.

It may also be possible to perform a direct reconstruction of $w_Q(z)$ and of
the quintessence potential $V$ from supernova observations \citep{HT99, CN00}.   
\cite{Saini00} have
attempted to perform reconstruction from the current supernova data and
find, unsurprisingly, that nothing definitive can be learned at present.
The theoretical studies that have been done have often encountered 
difficulty in accurately 
reconstructing the properties of a given model from simulated supernova
data \citep{HT00, WA01-2}.  This casts doubt upon the reliability
of reconstruction.  One of our main goals in the present study is 
to develop a reliable method for producing accurate reconstructions from
supernova data that is applicable to a wide class of quintessence models. 
Having done this, we will also want to explore whether
such reconstructions will be useful.  The method we use for reconstruction
demonstrates the power of using supernova observations to constrain 
cosmological parameters, and it highlights the usefulness of combining 
these observations with prior knowledge from other cosmological measurements.  
We therefore also undertake an exploration of the parameter constraints that
will be possible with SNAP and its combination with other experiments,
expanding on earlier analyses and discussing carefully what can and
cannot be learned in this way. 
 
This paper is arranged as follows. in Section~\ref{sec:quint} we summarize the 
cosmological effects of quintessence pertaining  to supernova observations.
In Section~\ref{sec:arbfit} we simulate SNAP-like data for a 	
quintessence-dominated universe and discuss the problems inherent in
producing reconstructions from such data. 
We discuss a cosmologically parameterized fitting function for $r(z)$
in Section~\ref{sec:cosfit}, and we use it for reconstruction in
Section~\ref{sec:cosrec}.    In Section~\ref{sec:parest} we show how more general 
questions about the nature of the dark energy may be addressed by constraining
parameters with SNIa data, and we discuss the long-term prospects 
for this sort of study in Section~\ref{sec:ltparam}.  We draw conclusions 
about the prospects for observational tests of dark energy in 
Section~\ref{sec:conclusion}

\section{Tracker Quintessence Cosmology}
\label{sec:quint}

We consider a scalar field (the quint\-essence field) $Q$, evolving in
a potential $V(Q)$.  The field will be nearly spatially
smooth, and for the purposes of this study, we ignore any small
inhomogeneities.  The density and pressure of 
quint\-essence are given by
\begin{eqnarray}
\label{eqn:quint_rho}
\rho_Q = {1\over2} \dot{Q}^2 + V(Q) \\
\label{eqn:quint_p}
p_Q = {1\over 2} \dot{Q}^2 - V(Q).
\end{eqnarray}
Quint\-essence is usually parameterized by its
equation of state $w_Q = p_Q / \rho_Q$.
 Looking at
equations~(\ref{eqn:quint_rho}) and~(\ref{eqn:quint_p}), we see immediately 
that $w_Q\geq -1$.
At late times, the coordinate distance to redshift $z$ is given by
\begin{equation}
r(z) = \int_0^z\frac{dz^\prime}
          {H_0\left[\Omega_M(1+z^\prime)^3 + \Omega_k (1+z^\prime)^2 + 
               \Omega_Q e^{\zeta(z^\prime)} \right]^{1/2}},
\label{eqn:cd_quint}
\end{equation}
where $\zeta(z) \equiv [3\int_0^z(1+w_Q(z^\prime))d\ln(1+z^\prime)]$.
In this paper we will assume a flat ($\Omega_k=0$) cosmology.
The coordinate distance (3) is of fundamental importance since
it fixes the magnitude-redshift relation probed by distant SNIa.

The evolution of quint\-essence models is governed by the
equation of motion
\begin{equation}
\label{eqn:motion}
\ddot{Q} + 3H\dot{Q} + V^\prime = 0,
\end{equation} 
where $V^\prime \equiv dV/dQ$.
One difficulty with quintessence is that, in most cases,
the initial conditions of this equation must be fine-tuned 
for quintessence to dominate only just at the present epoch, 
as is required by observations.  
It is possible to address this problem by using 
\emph{tracker potentials} \citep*{ZWS99,SWZ99}, which admit attractor-like 
solutions, thus allowing quintessence to exhibit identical behaviour
at late times for a wide range of initial conditions.
The energy scale of these models still requires fine-tuning to be
consistent with observations, however (see \cite{Vilenkin01} 
for further discussion of these issues).   The $k$-essence models 
\citep*{AMS00,AMS01} also address
the fine-tuning of initial conditions by adding to the 
quintessence Lagrangian kinetic terms that require quintessence to become 
dominant only after matter domination. Non-minimally coupled scalar fields 
offer another possible explanation of why quintessence domination might be related
to the epoch of radiation and matter equality (see \cite{Bean01} and 
references therein).  These models introduce extra
complications with little extra motivation, however, so to keep things 
simple we will concern ourselves with tracker models exclusively here.
We note that tracker models tend to evolve very slowly at late times, 
making it particularly difficult to detect evolution and reconstruct potental
shapes. So by focusing on tracker models here, we should be making a conservative
assessment of the prospects for testing quintessence models with supernova
observations.

Inverse power-law potentials $V\propto Q^{-P}$ provide the simplest tracker 
quintessence
models, but such models are inconsistent with the current observational
constraint $w_Q\la -0.6$ \citep*{PTW99,GPE99} for $P>2$.  
To keep things both simple
and realistic, then, we shall use a potential $V\propto Q^{-2}$ as our
standard test potential in this study.  [Recently it has 
been suggested that current observations may actually rule out
all inverse-power-law models \citep{CC01, BM02}, but they nevertheless
provide a useful theoretical test model.]  We shall see next
how a supernova measurement of $r(z)$ can be used to probe the properties of the 
dark energy.

\section{Re\-con\-struct\-ing the quintessence potential}
\label{sec:arbfit}

Various authors \citep{Starobinsky98, HT99, NC99} have derived equations 
for reconstructing the 
quintessence potential $V(Q)$ and equation of state $w_Q(z)$ 
from $r(z)$.  
The equation of state of the quint\-essence component is given by 
\begin{eqnarray}
\label{eqn:w_q_rec}
1+w_Q(z) &=& \frac{(1+z)}{3}\nonumber\\ 
    &\times&\frac{3\Omega_M(1+z)^2 + 
                  2 (d^2\tilde{r}/dz^2)/c^2 (d\tilde{r}/dz)^3}
         {\Omega_M(1+z)^3 - (c\, d\tilde{r}/dz)^{-2}},
\end{eqnarray}
and the quint\-essence potential is given parametrically by
\begin{eqnarray}
\label{eqn:Omega_q_rec}
\omega[\tilde{Q}(z)]&=& 
\left[\frac{1}{(d\tilde{r}/dz)^2} + \frac{(1+z)}{3}
       \frac{d^2\tilde{r}/dz^2}{(d\tilde{r}/dz)^3}\right]\nonumber \\
  &-& {1\over 2}\Omega_M(1+z)^3,
\end{eqnarray}
\begin{eqnarray}
\label{eqn:dqdz_rec}
\frac{d\tilde{Q}}{dz}& =& 
  \pm \frac{d\tilde{r}/dz}{(1+z)}\nonumber\\
 &\times&\left[-\frac{1}{4\pi}\frac{(1+z)d^2\tilde{r}/dz^2}{(d\tilde{r}/dz)^3}
          -\frac{3}{8\pi}\Omega_M(1+z)^3\right]^{1/2}.
\end{eqnarray}
Here we follow \cite{HT99} in using the dimensionless quantities
$\tilde{r} = H_0 r$, $\tilde{Q} = Q/M_{Pl}$, and
$\omega(\tilde{Q}) = V(Q)/\rho_{crit} = 
V(\tilde{Q}M_{Pl}) / (3H_0^2/8\pi G)$.  Thus, in principle, a
measurement of $r(z)$ will allow us to reconstruct the basic
properties of quintessence.

\begin{figure}
\begin{center}
\epsfig{figure = 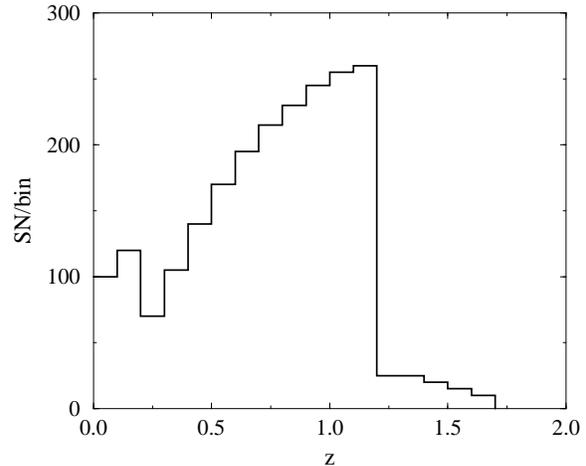, width=\linewidth}
\caption{The expected redshift distribution of supernovae
observations from the SNAP satellite \citep{SNAP}, augmented by ground-based 
low-redshift observations. This distribution is similar to that 
given in \citet{HT00}.}
\label{fig:snapzdist}
\end{center}
\end{figure}

To test reconstruction, we choose a tracker potential $V(Q)$,
evolve equation~(\ref{eqn:motion}) to obtain $w_Q(z)$ and $\Omega_Q$, 
and compute the Hubble-constant-free 
luminosity distance ${\mathcal{D}}_L(z) = (1+z) \tilde{r}(z)$ from 
equation~(\ref{eqn:cd_quint}). The magnitude-redshift relation
is then given by
\begin{equation}
\label{eqn:mz}
m(z) = {\mathcal{M}} + 5\log{{\mathcal{D}}_L(z)},
\end{equation}
where ${\mathcal{M}} = M-5\log{H_0}+25$ is the Hubble-constant-free
absolute magnitude, which we take to be ${\mathcal{M}}=-3.45$,
the best-fit value from \cite{GPE99}. From this, we draw
a Monte-Carlo sample of supernova magnitude-redshift datapoints
similar to what would be expected from the proposed SNAP
satellite~\citep{SNAP}.  Our simulated sample consists of 
roughly 2000 high-redshift supernovae from SNAP, in the range
$0.1\le z\le 1.7$, plus 200 low-redshift points from ground-based searches,
in the range $0<z\le 0.2$.  The points
have a measurement error of 0.01 magnitude, plus an intrinsic
scatter~\citep{GPEetal99} of 0.018 magnitude after correction for the decline
rate-luminosity correlation.  The error in 
the redshift is negligible, and the redshift distribution we use, 
(Fig.~\ref{fig:snapzdist}), is similar to the one used by \cite{HT00}.  
We neglect systematic errors in this analysis.

Given our set of N datapoints, we next choose a fitting function
$\tilde{r}^{\mathit{fit}}(z)$.  For a grid of points in the parameter 
space of $\tilde{r}^{\mathit{fit}}(z)$ we compute the likelihood function
\begin{equation}
\label{eqn:likelihood}
{\mathcal{L}} = \prod_{i=1}^N \frac{1}{\sqrt{2\pi\sigma_i^2}}
\exp{\left[-\frac{(m_i - m^{\mathit{fit}}(z_i))^2}{2\sigma_i^2}\right]},
\end{equation}
where $m_i$ is the magnitude of datapoint $i$, $z_i$ is its redshift, 
and $\sigma_i$ is 
the error.  The magnitude predicted by the fitting function
is given by $m^{\mathit{fit}}(z) = 
{\mathcal{M}} + 5\log{{\mathcal{D}}_L^{\mathit{fit}}(z)}$.
For the points in parameter space 
that fall within the $68\%$, $95\%$ and $99\%$ confidence
regions, we use $\tilde{r}^{\mathit{fit}}(z)$ to reconstruct 
$w_Q(z)$ and $V(Q)$ from 
equations~(\ref{eqn:w_q_rec}--\ref{eqn:dqdz_rec}).  We then draw confidence
contours around the outermost reconstructed curves from each confidence
region.

It is important to note here a basic mathematical difficulty with
reconstruction.  Looking at equation~(\ref{eqn:dqdz_rec}), we see
that reconstruction of $V(Q)$ is impossible when the expression in 
square brackets crosses zero.  When this occurs at a given confidence
level, we terminate the corresponding reconstruction contour for $V(Q)$.
Also, we note that the present-day value of the quintessence field $Q$ is not
fixed by equation~(\ref{eqn:dqdz_rec}).  But since this value has no
cosmological effect, we may arbitrarily set $Q=1$ at the present day.

\begin{figure*}
\begin{center}
\epsfig{figure=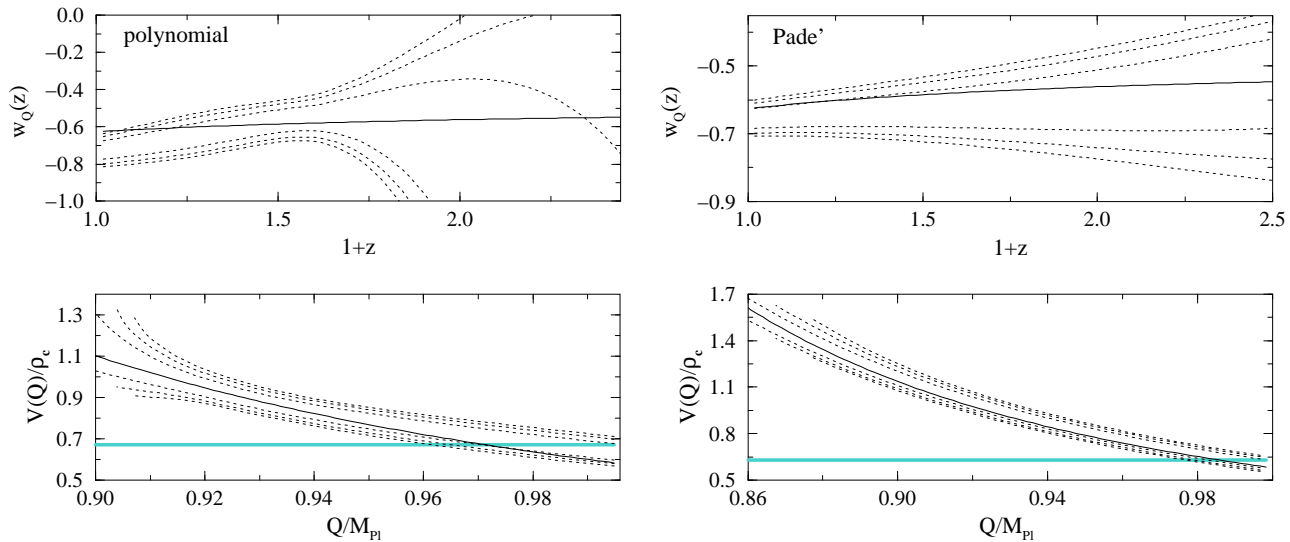, width=\linewidth}
\caption{Reconstruction of $V(Q)$ and
$w_Q(z)$ for a quint\-essence model with $V\propto Q^{-2}$. 
In the left-hand panels we have used the third-order 
polynomial~(\ref{eqn:polyfit}) as a fitting function, and
in the right-hand panels we have used the Pad\'{e}
approximate~(\ref{eqn:padefit}).
The dashed lines---from innermost to outermost pairs---are 
$68\%$, $95\%$ and $99\%$ confidence contours,
and the solid black lines are the actual model values.  
The heavy lines in the lower panels are lines of constant $V$, 
for reference. Inaccuracies in the derivatives
of the fitting functions lead to inaccurate reconstructions
of $w_Q(z)$ and, for the polynomial fit, of $V(Q)$.}
\label{fig:badrec}
\end{center}
\end{figure*}

Various functional forms for $\tilde{r}^{\mathit{fit}}(z)$ have been
recommended in the literature for use in reconstruction 
\citep{HT00,WA01-2,CN00,Saini00}. To show the difficulties
inherent in reconstruction, we attempt reconstructions using 
two of them---a third order polynomial
\begin{equation}
\tilde{r}^{\mathit{fit}}(z) = a_3 z^3 + a_2 z^2 + a_1 z,
\label{eqn:polyfit}
\end{equation}
and a Pad\'{e} approximate
\begin{equation}
\label{eqn:padefit}
\tilde{r}^{\mathit{fit}}(z) = \frac{z(1+az)}{1+bz+cz^2},
\end{equation}
both of which have been suggested by \cite{HT99,HT00}.
Reconstructions resulting from fitting SNAP-like data with
these fitting functions are shown in Fig~\ref{fig:badrec}.
In applying the reconstruction equations here, we have assumed that 
$\Omega_M$ is known exactly.  Evidently, the precision of the 
reconstructions will diminish if we allow for some uncertainty
in $\Omega_M$. 

The reconstructions we obtain are inaccurate. 
For the polynomial fit,
the actual value of $w_Q$ lies outside the $99\%$ confidence contour
at low redshift, where the reconstruction is most precise, and
the actual value of $V(Q)$ lies outside the $68\%$ contour at
high values of $Q$.  The Pad\'{e} fit does significantly better, but
the reconstruction of $w_Q(z)$ is still slightly inaccurate at low $z$.
Reconstructions based on these fitting functions thus appear to be unreliable.
Since our reconstruction contours rule out the \emph{correct} model,
we cannot seriously claim that they rule out \emph{any} specific model:
they are too inaccurate to be useful.  (Similar difficulties
with the 3rd-order polynomial fit were noted by
\cite{WA01-2} in a preprint which appeared as this work was being completed.)

\begin{figure}
\begin{center}
\epsfig{figure=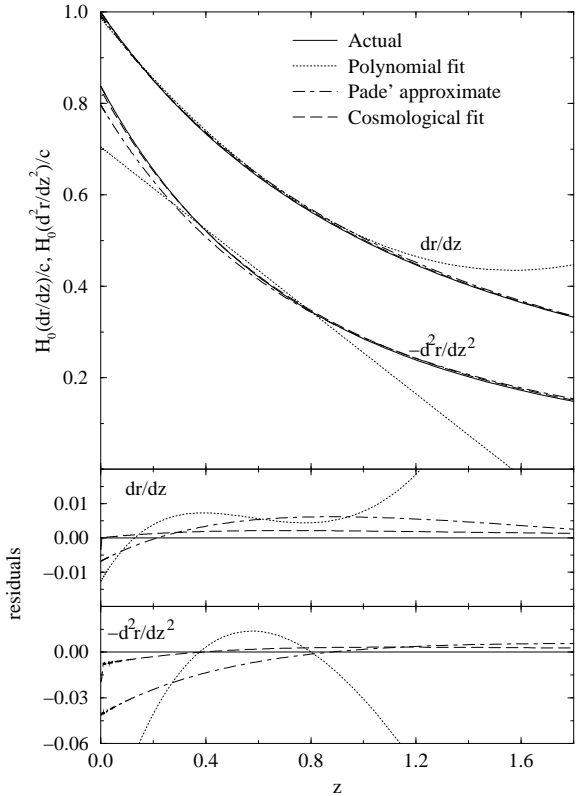, width = \linewidth}
\caption[Comparison of derivatives of $r(z)$ with derivatives
of fitting functions]{Derivatives of the best fits to $r(z)$ for various 
fitting functions, compared with the actual derivatives 
$dr/dz$ and $d^2r/dz^2$ (solid lines).  
The third-order polynomial fit [equation~(\ref{eqn:polyfit}), dotted lines],  
despite producing a good fit to $r(z)$, gives an obviously 
poor fit to the derivatives. 
It will therefore produce poor reconstructions of $V(Q)$ and $w(z)$.
The Pad\'{e} approximate fit  [equation~(\ref{eqn:padefit}), 
dot-dashed lines] provides more reasonable derivatives, but 
the small discrepancies between the fit and the actual 
derivatives still lead to inaccuracies in reconstruction.
The fit based on a  cosmological parameterization 
[equation~(\ref{eqn:fit_wln}), dashed lines] 
provides the best fits to $dr/dz$ and $d^2r/dz^2$,
as may be expected, since the fitting function here is an approximation
to the actual coordinate distance. } 
\label{fig:derivfig}
\end{center}
\end{figure}

This difficulty arises because the reconstruction 
equations~(\ref{eqn:w_q_rec}--\ref{eqn:dqdz_rec}) depend on the
derivatives of $\tilde{r}(z)$, rather than on $\tilde{r}(z)$ itself,
but the derivatives of the fitting functions do not necessarily
resemble the derivatives of $\tilde{r(z)}$. 
(This problem has also been noted by \cite{HT00} and \cite{WA01-2}.) 
The problem with the fitting functions suggested in the literature is that, 
with the  exception of the one
in \cite{Saini00}, they are all arbitrary functions that have
been chosen because they give good fits to $\tilde{r}(z)$.  
However, we have no \emph{a priori} reason to believe that their derivatives
will resemble $d\tilde{r}/dz$ and $d^2\tilde{r}/dz^2$. Indeed,
as Fig.~\ref{fig:derivfig} shows, the best fitting polynomial and
Pad\'{e} approximate do not provide particularly accurate fits to
the derivatives of $r(z)$.  Therefore we should not expect to obtain 
accurate reconstructions from these functions.
To perform reconstructions reliably, we must first find a fitting function 
that we believe
will provide not only a good fit to $\tilde{r}(z)$, but also a good
approximation to its derivatives.  In essence, we want to fit our data 
with some physically motivated approximation to $\tilde{r}(z)$.

\section{A physically motivated fitting function}
\label{sec:cosfit}

We can construct a physically motivated fitting function by noting that,
for low $z$, the quint\-essence equation of state is often well 
approximated by
\begin{equation}
\label{eqn:w_q_app}
w_Q^{app} = w_Q(a_0) + \alpha \ln(a/a_0)
          = w_Q(z=0) - \alpha\ln(1+z),
\end{equation}
where $\alpha \equiv dw_Q / d\ln a$ \citep{GPE99}.  
Combining this with equation~(\ref{eqn:cd_quint}), we arrive an at 
approximate expression for the coordinate distance, which we 
can use as a fitting function for reconstruction:
\begin{equation}
\label{eqn:fit_wln}
\tilde{r}^{\mathit{fit}}(z) =  
\int_0^z{ \frac{dz^\prime}
              {\left[\Omega_M(1+z^\prime)^3 +
                      \Omega_Q^{app}(z^\prime)
                       \right]^{1/2}}}, 
\end{equation}
where the approximate energy density contribution of the quint\-essence 
component is
\begin{equation}
\Omega_Q^{app}(z) = 
\Omega_Q(1+z)^{3(1+w_Q(0))} 
               \exp\{{-(3/2)\alpha[\ln(1+z)]^2\}}.
\end{equation} 
Also, by assumption, $\Omega_M = 1-\Omega_Q$ (present-day values). 
Equation~(\ref{eqn:fit_wln}) is readily twice differentiable with
respect to $z$; hence it can be used to reconstruct $V(Q)$ and 
$w_Q(z)$. A similar fitting function [with a linear, rather than
logarithmic, approximation to 
$w_Q(z)$]  has been found to provide the best fits to $r(z)$ of any of 
the fitting functions previously considered in the literature \citep{WA01}. 
Fig.~\ref{fig:derivfig} shows that equation~(\ref{eqn:fit_wln})
also produces better fits to $d\tilde{r}/dz$ and 
$d^2\tilde{r}/dz^2$ than the polynomial or
Pad\'{e} approximate, suggesting that it may produce reliable 
reconstructions.  This is entirely to be expected, since the fitting 
function is a direct approximation to the coordinate distance
in terms of real physical parameters.  Of course this function will
produce  reliable reconstructions only to the extent that it is a good
approximation to $\tilde{r}(z)$,  but we expect that it will 
approximate $\tilde{r}(z)$ reasonably well, for a wide
range of quint\-essence models, out to at least the redshifts to be probed by 
SNAP \citep{GPE99}.  It is worth noting that some models
will not be well approximated by equation~(\ref{eqn:fit_wln})---for example, the
pseudo-Nambu-Goldstone boson potentials considered by \cite{WA01-2} (which can produce
an oscillatory equation of state) would generally require a
 more complicated fitting function.  Nevertheless,
the analysis presented here is self-consistent: as long as 
equation~(\ref{eqn:fit_wln}) provides an acceptable fit to the data (reduced $\chi^2$
of order unity), we are unlikely
to learn anything more by introducing more complicated fitting functions.  If there
are significant residuals about the best fit, however, it may be fruitful to consider
more complicated functions.

It is also important to note that the use of equation~(\ref{eqn:fit_wln})
as a fitting function effectively reduces the problem of reconstruction to 
one of constraining the cosmological parameters $\Omega_Q$, 
$w_Q(0)$ and $\alpha$ by likelihood analysis. Parameter estimation and reconstruction
are separately interesting: the former can tell us about the basic properties of
dark energy in a model-independent way, whereas the latter can tell us more
specifically about the potential of a scalar-field quintessence.
Since we are constraining
some cosmological parameters in the reconstructions discussed here, it may
be possible to combine supernova observations with other observations 
to better constrain the fit parameters and simultaneously to improve our 
reconstructions. For example, in a flat universe (which we assume), measurements of 
$\Omega_M$ from, \emph{e.g.}, galaxy clustering and the CMB fix $\Omega_Q$
via the constraint equation  $\Omega_Q=1-\Omega_M$. 
If there are degeneracies between the parameters, then
combining such measurements with supernova data will serve to tighten
the constraints on all the parameters, improving the precision
of our reconstructions.  Because we have used physical quantities to
parameterize our fitting function for the coordinate distance, we can learn about
$\tilde{r}^{\mathit{fit}}(z)$ indirectly from data other than the SN1a observations.  
In this way, a 
physically motivated fitting function allows reconstructions that are not only
more accurate but also possibly more precise than reconstructions from more arbitrary
fitting functions.

\begin{figure*}
\begin{center}
\epsfig{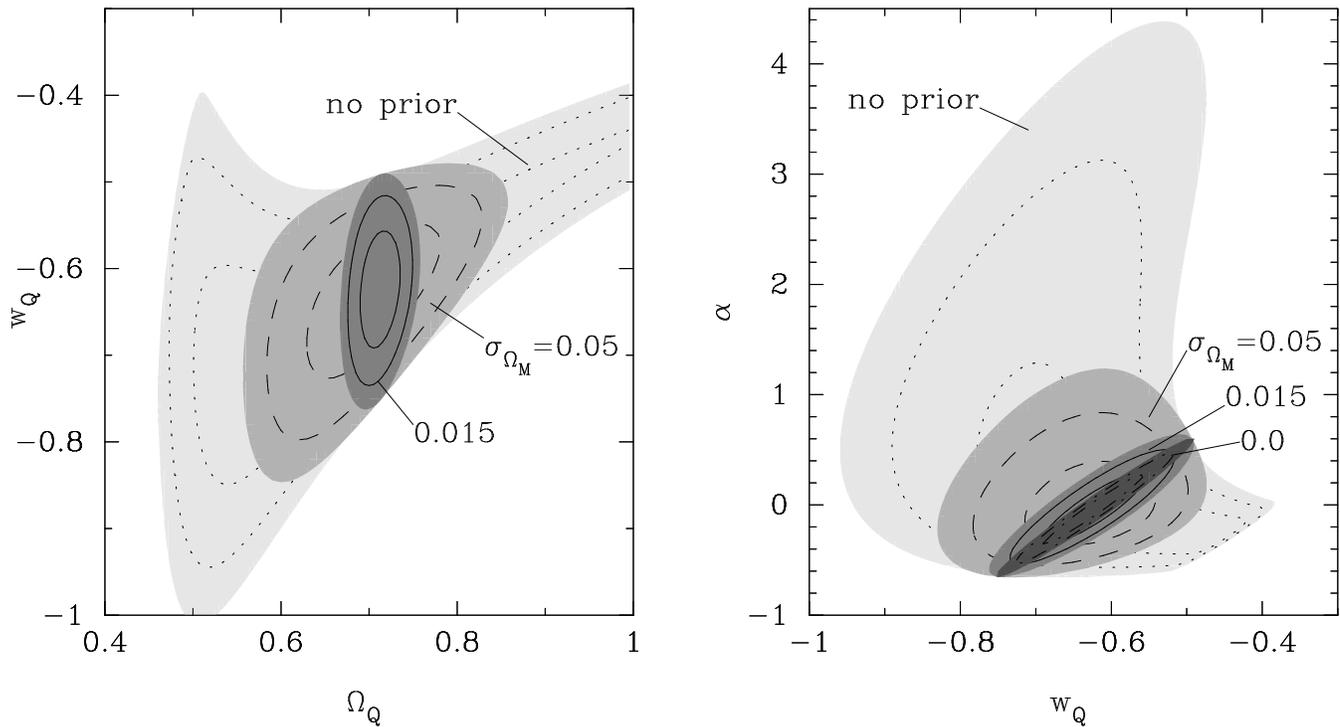}
\caption[Likelihood contours for $\Omega_Q$, $w_Q(0)$ and $\alpha$, with
prior constraints on $\Omega_M$]{Likelihood contours for the 
cosmological parameters
$\Omega_Q$, $w_Q(0)$ and $\alpha$, as defined in 
equation~(\ref{eqn:fit_wln}).  The contours are calculated from 
a SNAP-like dataset for one year of observation, 
drawn from a quint\-essence model with
$V(Q)\propto Q^{-2}$.  The shaded regions
indicate $99\%$ confidence regions, given various Gaussian prior
probability distributions on $\Omega_M=1-\Omega_Q$, with standard
deviation $\sigma_{\Omega_M}$.  
The contour lines inside each shaded region indicate the associated 
$95\%$ and $68\%$ confidence
levels.  Note that strong degeneracies between parameters make
precise parameter estimation impossible in the absence of prior knowledge.
In each graph, the two-dimensional
likelihood contours shown are marginalized over the third parameter.}
\label{fig:contour_P2}
\end{center}
\end{figure*}

Fig.~\ref{fig:contour_P2} shows the sort of constraints on
$\Omega_Q$, $w_Q(0)$ and $\alpha$ 
that might be expected from one year of SNAP data (2200 SN).  
Clearly, strong and complicated degeneracies between the
parameters make precise constraints impossible when
we use the supernova data alone for estimation.  
It may be possible to reduce these degeneracies 
significantly if we combine the supernova data with other
measurements of the cosmological parameters.  For example, galaxy clustering
measurements from the 2dF galaxy redshift survey, combined with measurements of
CMB anisotropies, constrain $\Omega_M$ to within roughly $\pm 0.1$ ($2\sigma$ errors)
in the case of a flat universe \citep{GPEetal02}. 
To show the effect that such other measurements can have, we impose a Gaussian prior
probability distribution on $\Omega_M = 1-\Omega_Q$. Its effect on
the contours, for
various values of the standard deviation $\sigma_{\Omega_M}$, is shown in 
Fig.~\ref{fig:contour_P2}.  Clearly the use of such a prior
is significant:  if we impose 
$\sigma_{\Omega_M} = 0.05$, it is possible to constrain
$w_Q(0)\neq -1$ with $99\%$ confidence, indicating that the 
dark energy is not a cosmological constant.
Other authors \citep{Astier00,BM00,Goliath01, HT00,MBS01,WA01,WA01-2} 
have attempted to constrain a similar set of parameters 
using a SNAP-like dataset (using 
$w_1\equiv dw_Q/dz$ where we have used $\alpha\equiv d\ln{w_Q}/d\ln{a}$).
They also find that strong degeneracies between parameters
make precise estimation impossible, unless prior knowledge of one
of the parameters is assumed.  Our results are in good agreement with
these analyses.\footnote{Recently, other authors \citep{WG01, WL01} have
suggested, interestingly, that better information about dark energy may be obtained
by constraining the function $f(z)=e^{\zeta(z)}$ (where $\zeta(z)$ is as defined after 
equation~(\ref{eqn:cd_quint})) at various values of $z$, rather than by attempting
to parameterize and constrain $w(z)$.}

But is our Gaussian prior with $\sigma_{\Omega_M} = 0.05$ realistic?  
The prior assumes that we will be able to measure $\Omega_M$ to this
accuracy independent of the equation of state, that is, that the
measurement of $\Omega_M$ does not suffer from degeneracies with $w_Q$.  
In fact, measurements of $\Omega_M h$ from the 
Sloan Digital Sky Survey (SDSS), combined with measurements of
$\Omega_M h^2$ from the MAP satellite's CMB measurements will be 
able to constrain $\Omega_M$ to nearly $\pm0.05$, with no dependence 
on $w_Q$ \citep{EHT98}.
Moreover, if flatness is assumed and this measurement is combined with 
measurements of the angular diameter distance to last scattering from the
the MAP or Planck satellites, the uncertainty in $\Omega_M$ decreases
to $0.01$--$0.03$, with an additional strong constraint on $w_Q$ and
no degeneracy between the parameters \citep{Huetal98}. In addition,
proposed X-ray or Sunyaev-Zeldovich effect surveys of galaxy clusters
can constrain $\Omega_M$ to within an error of $0.05$ or better, 
with only a very slight dependence on $w_Q$ \citep*{HMH01, WBK01}.  A proposed 
gravitational lensing survey could constrain $\Omega_M$ to within $0.015$ 
with minimal $w_Q$ dependence \citep*{WBM99}.
  
The uncertainty $\sigma_{\Omega_M} = 0.05$ thus appears 
to be a conservative estimate of the precision that may be obtained in 
the future.  More optimistically, we could use a prior with 
$\sigma_{\Omega_M} = 0.015$.  Fig.~\ref{fig:contour_P2} shows the
constraints that are possible in this case: with improved knowledge of 
$\Omega_M$ we can also obtain much tighter constraints on $w_Q(0)$
and $\alpha$.  In no case is it possible to show that $\alpha\neq 0$, 
however, even if $\Omega_M$ is known exactly.  Nevertheless, it is clear
that combining SNAP parameter constraints with other measurements will
be a powerful probe of the dark energy.  We shall explore this in more
detail below.

\section{Reconstruction from Cosmologically Motivated Fitting Functions}
\label{sec:cosrec}

\begin{figure}
\begin{center}
\epsfig{figure=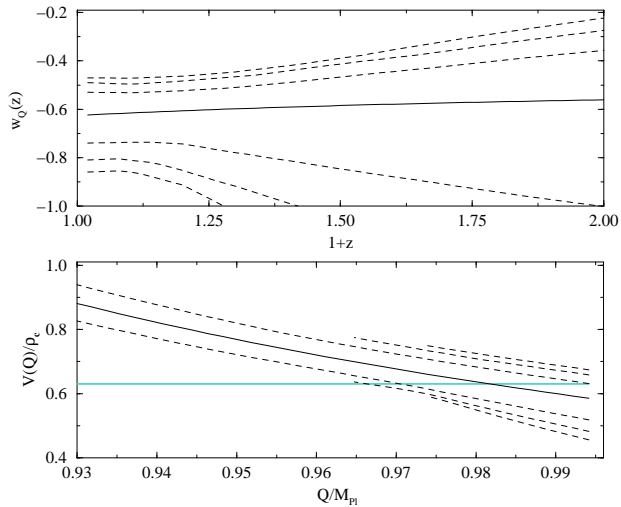, width=\linewidth}
\caption[Reconstruction from cosmological fitting function for 
$V\propto Q^{-2}$, with $\sigma_{\Omega_M} = 0.05$ ]{Reconstruction 
of $V(Q)$ and $w_Q(z)$ for a quint\-essence
model with $V\propto Q^{-2}$, using
equation~(\ref{eqn:fit_wln}) as a fitting function.  A Gaussian
prior is imposed on $\Omega_M$, with standard deviation
$\sigma_{\Omega_M} = 0.05$ about the actual value.
Both $w_Q$ and $V$ are accurately constrained, and we are able
to constrain $V(Q)\neq const$ at $68\%$ confidence.}
\label{fig:wlnrec_P2_sig.05}
\end{center}
\end{figure}

Prior knowledge of the parameters is also necessary if we 
are to use equation~(\ref{eqn:fit_wln}) successfully for reconstruction.
If we impose no prior probability distribution on our likelihood analysis
(the ``no prior'' case in Fig.~\ref{fig:contour_P2}), 
then reconstruction of $V(Q)$ fails for nearly all $Q$ at all confidence
levels, and $w_Q(z)$ is only very weakly constrained. If, on the other 
hand, we impose a Gaussian prior with $\sigma_{\Omega_M} = 0.05$,
we obtain the potentially useful reconstructions shown in  
Fig.~\ref{fig:wlnrec_P2_sig.05}. 
(Note that because we are explicitly constraining
$\Omega_Q$---and hence $\Omega_M$, which appears explicitly in the
reconstruction equations---we have allowed $\Omega_M$ to vary over
its allowed range in applying the reconstruction equations.)  In particular,
$V(Q)$ is constrained to be non-constant at the $68\%$ confidence
level.   More importantly,the reconstructions are accurate:  
both the actual $V(Q)$ and the actual $w_Q(z)$ curves are contained 
within the $68\%$ confidence contours.  This is very promising: it should
be possible, by combining SNAP data with expected measurements of 
$\Omega_M$, to reconstruct the potential of an inverse-power-law
quintessence model in an accurate and reliable way. 

\begin{figure}
\begin{center}
\epsfig{figure=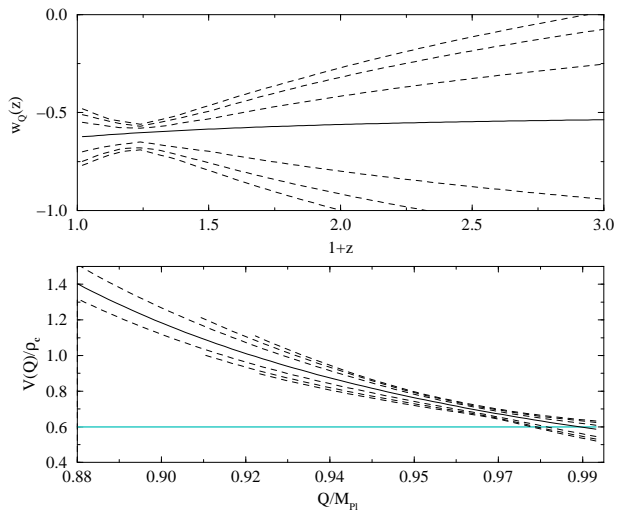, width=\linewidth}
\caption[Reconstruction from cosmological fitting function for 
$V\propto Q^{-2}$, with $\sigma_{\Omega_M} = 0.015$]{Similar to 
Fig.~\ref{fig:wlnrec_P2_sig.05}, except here
$\sigma_{\Omega_M} = 0.015$.  $V(Q)$ is constrained to be non-constant
at $>99\%$ confidence and $w_Q(z)$ is constrained
with good precision.}
\label{fig:wlnrec_P2_sig.015}
\end{center}
\end{figure}

\begin{figure}
\begin{center}
\epsfig{figure=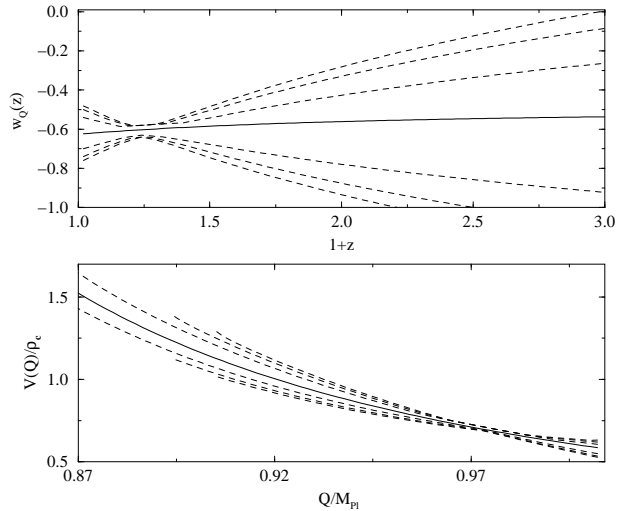, width=\linewidth}
\caption[Reconstruction from cosmological fitting function for 
$V\propto Q^{-2}$, with $\Omega_M$ fixed]{Similar to 
Fig.~\ref{fig:wlnrec_P2_sig.05}, except here
we hold $\Omega_M$ fixed at its true value, rather than allowing it 
to vary.  The fact that the reconstructions are accurate in this case
indicates clearly that the cosmological fitting 
function~(\ref{eqn:fit_wln}) is a better 
fitting function for reconstruction than the arbitrary ones we tried
previously. }
\label{fig:wlnrec_P2_Omexact}
\end{center}
\end{figure}

Fig.~\ref{fig:wlnrec_P2_sig.015} shows
a reconstruction using the more optimistic prior $\sigma_{\Omega_M}=0.015$. 
In this case, we are able to constrain $V(Q)$ to be non-constant
at greater than $99\%$ confidence with good accuracy, 
and we achieve impressive constraints
on $w_Q(z)$, although it is not possible to show that $w_Q(z)$ is 
evolving.  Because the reconstructions we obtain are accurate, it appears 
that equation~(\ref{eqn:fit_wln}) is a better
fitting function for reconstruction than the arbitrary fitting
functions we tried previously.  But we cannot be entirely sure
of this yet:  when we attempted reconstructions with the polynomial
and Pad\'{e} approximate, we assumed exact knowledge
of $\Omega_M$.  
It is possible that our allowing $\Omega_M$ to vary here might
be obscuring inaccuracies.  To test this, in 
Fig.~\ref{fig:wlnrec_P2_Omexact} we perform a reconstruction
from equation~(\ref{eqn:fit_wln}), in which we hold $\Omega_M$ 
(and thus $\Omega_Q$) constant at its true value throughout. 
The reconstructions are entirely acceptable: no obvious inaccuracies 
creep in when we fix $\Omega_M$. As hoped, the use of a 
physically motivated fitting function has given us a method for
reconstructing the properties of quintessence in an accurate
and reliable way.

We note that \cite{HT00}
(HT00) have previously attempted to constrain $w_Q(z)$ using a fitting
function based on $w_Q^{app}(z)$. Our work differs in  several important 
respects from their analysis.  First, we show here that 
equation~(\ref{eqn:fit_wln}) produces good reconstructions of $V(Q)$ 
as well as $w_Q(z)$.  Second, whereas HT00 hold $\Omega_M$ fixed in their analysis, 
we show that good reconstructions are still possible when there is some uncertainty
in $\Omega_M$.  Third, HT00 use the Fisher matrix to draw their 
confidence contours, while we use an exact likelihood analysis.
 Most importantly, HT00 draw their simulated data 
from a model with $w_Q$ exactly equal to $w_Q^{app}$, which guarantees an 
accurate reconstruction (since the fitting function is the same as the 
original model).  We show here that accurate reconstructions can be done 
even when $w_Q^{app}$ is only an approximation to the actual equation
of state.  \cite{WA01-2} have recently had some success in reconstructing 
$w_Q(z)$ using a polynomial expansion of $w_Q(z)$ to produce a fitting 
function, although they have also held $\Omega_M$ fixed.  
Also, since the equation of state is not usually well
approximated by a linear function of $z$, they must expand $w_Q$ to second order to
obtain accurate reconstructions, which requires one more fit parameter
than our fitting function. We expect that this will lead to worse 
degeneracies in parameter space 
and that it may be difficult to obtain a useful reconstruction if $\Omega_M$ 
is allowed to vary.

So it appears that, by fitting our data to equation~(\ref{eqn:fit_wln}), 
we have found a way to reconstruct $V(Q)$ and 
$w_Q(z)$ in a reliable, accurate and fairly precise way.
Recall, however, that, in deriving equation~(\ref{eqn:fit_wln}), 
our basic approximation was to write the equation of state in the form 
$w_Q^{app}(z)=  w_Q(0) - \alpha\ln(1+z)$.
So what we are actually reconstructing when we use
equation~(\ref{eqn:fit_wln}) as a fitting function is the
$w_Q(z)$ that \emph{would be} most likely to produce the data
if $w_Q(z)$ had the form of $w_Q^{app}(z)$.  Likewise, the
$V(Q)$ that we reconstruct is the potential that \emph{would have} 
produced this most likely $w_Q^{app}(z)$. 
If we are careful to take these issues into account, it 
will be possible to make use of our reconstructions to distinguish
between different quintessence models.  As long as $w_Q(z)$ is well
approximated by $w_Q^{app}(z)$ over the redshifts
we are measuring (\emph{i.e.}, as long as it is not dominated
by nonlinear terms in $\ln[1+z]$), we expect that our reconstructions 
will be accurate.  Clearly models with rapidly varying $w_Q(z)$ (\emph{e.g.}, the
pseudo-Nambu-Goldstone boson models discussed by \cite{WA01-2}) will not
be accurately reconstructed using equation~(\ref{eqn:fit_wln}), but in these 
cases we also would not expect an acceptable fit to the data. 

If we want to compare a given potential $V(Q)$ to our reconstructions, 
we must first determine whether the $w_Q(z)$ arising from
it is well approximated by $w_Q^{app}(z)$.  If it is not well approximated, 
we can say nothing useful about that particular potential using our 
reconstructions, since we cannot be sure that reconstructions of it 
would be accurate. If, on the other hand, the model in question is well
approximated by  $w_Q^{app}(z)$, we can determine whether it is
allowed by the data. First, we vary the parameters of the
potential $V(Q)$, including the present-day value of the field 
$Q_0$, since this is not fixed by the reconstruction
equations.  If we do not find any parameterization that falls entirely 
within the reconstruction contours, then the model under consideration
is disallowed by the data.  If, on the other hand, we do find a satisfactory
parameterization, we must then additionally check whether or not the 
resulting equation of state falls  within the reconstruction contours
for $w_Q(z)$ for the redshift range of the data.  If it does not,
then the model is ruled out.  Any model that passes both of these
tests is consistent with the data.

\begin{figure}
\begin{center}
\epsfig{figure=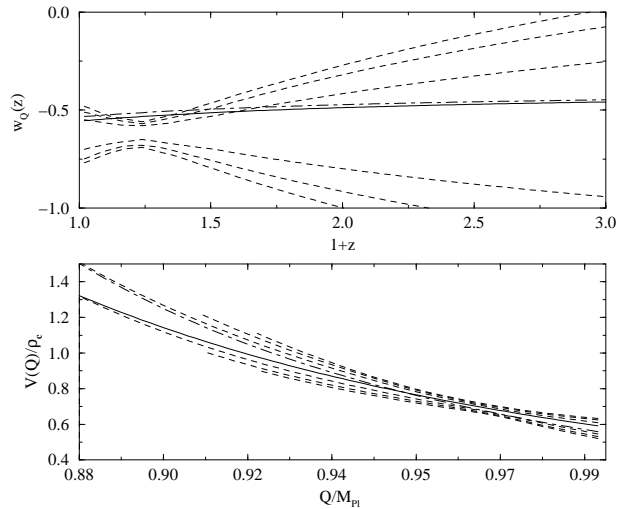, width=\linewidth}
\caption[]{Comparison of models with  $V(Q) = M^7/ Q^{3}$ 
to the confidence contorus in Fig.~\ref{fig:wlnrec_P2_sig.015}.
In the lower panel, the dot-dashed line and solid line correspond, 
respectively, to the minimum and maximum values of $M$ for which 
the potential falls between the $68\%$ confidence contours.  The
corresponding lines in the upper panel are $w_Q(z)$ that arise from evolving 
quintessence in these potentials.  Because $w_Q$ decreases monotonically
with increasing $M$ for inverse-power-law models, all allowed values
of $w_Q(z)$ fall between these two lines.  Clearly, models with
$V\propto Q^{-3}$ are ruled out by the data.   }
\label{fig:extreme_pars}
\end{center}
\end{figure}

As an example, we test an inverse-power-law model with $V\propto Q^{-3}$
against the contours shown in Fig.~\ref{fig:wlnrec_P2_sig.015}
(recall that these arose from a quintessence model with $V\propto Q^{-2}$).
We vary the parameters $M$ and $Q_0$ for the potential
$V(Q)= M^7/[Q-(1-Q_0)]^3$ and identify
a range of $M$ and $Q_0$ for which the potential falls within the
$68\%$ confidence contours.  This gives us a minimum and maximum
allowed value for $M$.  We then evolve the quintessence field
in the potential $V(Q) = M^7/Q^3$, to get $w_Q(z)$ curves for both of these 
values. (The transformation $Q\rightarrow Q-(1-Q_0)$
has no cosmological effect.)  For 
inverse-power-law models $w_Q$ decreases monotonically with increasing
$M$; therefore, all allowed potentials will have $w_Q(z)$ values
that fall between these two limiting curves.  As Fig.~\ref{fig:extreme_pars} shows,
$w_Q(z)$ falls outside the $99\%$ confidence contours for all allowed
values of $M$, and so models with $V\propto Q^{-3}$ are ruled out
by the data.  The reason that the allowed potentials do not lead to
allowed $w_Q(z)$ curves here is that the calculation is not self-consistent.
Each allowed parameterization of $V(Q)$ includes a specific value for
both $M$ and $Q_0$, but only $M$ has any effect on the evolution of the
quintessence field.  For each allowed value of $M$, if the quintessence field
does not evolve to the associated $Q_0$ at the present epoch, the resulting
$w_Q(z)$ may still be disallowed.

Reconstructions using the fitting function equation~(\ref{eqn:fit_wln})
are thus clearly more useful than reconstructions using arbitrary
fitting functions.  Using this equation, it is possible to 
determine, in a consistent way,  whether or not a particular model
is ruled out by the data---so long as $w_Q^{app}(z)$
is a good approximation to $w_Q(z)$ for that model. With the polynomial and
Pad\'{e} approximate fitting functions we tried before, we would
have no reliable way of ruling out specific models, since we would
have no good way of determining whether the fitting function accurately
approximates the derivatives of $\tilde{r}$.
Any apparent conclusion from arbitrary fitting functions like these 
could be spurious, the result of unaccounted-for
inaccuracies in the fitting function or its derivatives.
If we wish to produce useful reconstructions, it will be necessary
to use a fitting function similar to equation~(\ref{eqn:fit_wln}).

Testing individual models as described above could be useful, but because
we lack any good theoretical motivation for model-building at present, testing 
the large numbers of possible models will be very time-consuming.  
For this reason, it may
be more worthwhile to consider more general questions about the dark 
energy via likelihood analysis in parameter space, 
rather than to test models individually via reconstruction.

\section{Parameter estimation: determining the properties of dark energy}
\label{sec:parest}

There are two especially important questions we would like to address
concerning the properties of the dark energy.  First, we would like
to know whether the equation of state $w_Q$ differs from $-1$---that is, 
whether the dark energy is the cosmological constant or not---and second, 
we would like to know whether $w_Q$ varies with redshift.  In the previous
section, we saw that it is possible to  address these
questions by performing a likelihood analysis over the parameters
$w_Q(0)$, $\alpha$ and $\Omega_Q$. In Fig.~\ref{fig:contour_P2},
we saw that it is possible, for an inverse-power-law quintessence model, 
to constrain $w_Q(0) \neq -1$ at $95\%$ confidence using SNAP
data alone.  If we also include prior knowledge of $\Omega_Q$, then
it is possible to rule out $\Lambda$ models at even higher confidence.
Even with exact knowledge of $\Omega_Q$, though, it is not possible
to rule out $\alpha = 0$ for this model; it is not possible in 
this case to say whether $w_Q$ is changing with redshift. This result
is in agreement with several previous studies \citep{Astier00,BM00,Goliath01, 
HT00, MBS01,WA01, WA01-2}.

\subsection{Two-parameter fits}
\label{sec:like_2p}

It is also clear from Fig.~\ref{fig:contour_P2} that strong degeneracies 
between $\alpha$ and the other parameters lead to large uncertainties in 
our parameter estimation. Since we have not been able to say anything
particularly useful about $\alpha$, it might be more useful to
leave $\alpha$ out of the analysis entirely for the sake of parameter
estimation, focusing our attention on the value of $w_Q$ exclusively.
To do this, we derive a  fitting function for $\tilde{r}(z)$
based on an effective \emph{constant} equation of state 
$w_Q^{\mathit{eff}}$, 
\begin{equation}
\tilde{r}^{\mathit{fit}}(z) = 
\int_0^z\frac{dz^\prime}{\left[\Omega_M(1+z^\prime)^3 + 
                \Omega_Q (1+z^\prime)^{3(1+w_Q^{\mathit{eff}})}\right]^{1/2}} .
\label{eqn:rfit_2par}
\end{equation}
Several other authors have considered this parameterization of
the coordinate distance, and
current supernova observations constrain this parameter to be
$w_Q^{\mathit{eff}}\la -0.4$ \citep{SCP}. When these observations are combined with
CMB \citep{GPE99} or large-scale structure \citep{PTW99} data, the 
constraint becomes $w_Q^{\mathit{eff}}\la -0.6$. Recently there has been 
some suggestion that CMB data slightly favors $w_Q^{\mathit{eff}}>-1$ 
\citep{Baccigalupi01}, 
although this conclusion assumes that some rather poorly constrained parameters 
are known exactly.

\begin{figure}
\begin{center}
\epsfig{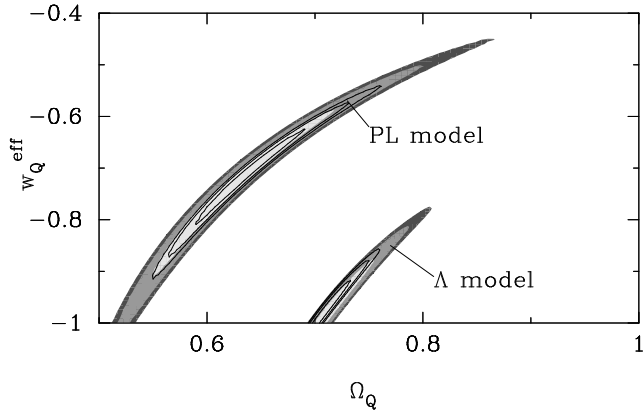}
\caption[Likelihood contours for $\Omega_Q$ and $w_Q^{\mathit{eff}}$,
for $\Lambda$ and quintessence]{Likelihood contours for $\Omega_Q$ and 
the effective constant
equation of state $w_Q^{\mathit{eff}}$ from simulated SNAP data.  The contours
shown are for an inverse-power-law quintessence model with 
$V(Q)\propto Q^{-2}$ (``PL model'') and for a cosmological constant
model (``$\Lambda$ model'').  Shaded contours are $68\%$, $95\%$ and $99\%$
confidence contours for one year of SNAP
data (2200 SNe). The line contours define the same confidence regions
for three years of data (6600 SNe).
Clearly, it is possible to distinguish the quintessence model from
a $\Lambda$ model with high confidence for a full three-year SNAP dataset.}
\label{fig:2pcont}
\end{center}
\end{figure}

Fig.~\ref{fig:2pcont} shows the likelihood contours that are obtained
in the parameter space of equation~(\ref{eqn:rfit_2par}) for SNAP-like
datasets drawn from a  $\Lambda$ model and a tracker model with potential 
$V(Q)\propto Q^{-2}$.  The figure shows the confidence contours that might
be expected for one year and three years of SNAP data.  No priors have 
been imposed on either of the parameters.  
In the case of quintessence, with one year of 
data it is possible to constrain $w_Q^{\mathit{eff}} \neq -1$ with
$68\%$ confidence, and with three years of data it is possible to do
so with $99\%$ confidence.  Since $w_Q = -1$ for a $\Lambda$ model,
a cosmological constant is ruled out at the same confidence levels.
The fact that we can use SNAP to distinguish between a cosmological
constant and quintessence is very encouraging.  So far, however, we have
only shown this for the specific case of an inverse-power-law
quintessence model.  It is not too surprising that $w_Q$ can be
distinguished from $-1$ in this case, since this model has present-day
equation of state $w_Q(0)\approx -0.6$, relatively far from $-1$, when 
$\Omega_Q \approx 0.7$.
It will be interesting to see if we can obtain similar results for a 
broader range of quintessence models---especially those with lower
values of $w_Q$.

To test this, we consider two different tracker potentials.  
The first is a potential inspired by supergravity \citep{SUGRA},
\begin{equation}
V(Q) = \frac{M^{4+P}}{Q^P}
\exp\left[\frac{1}{2}\left(\frac{Q}{M_{Pl}}\right)^2\right].
\label{eqn:SUGRA}
\end{equation} 
This potential exhibits tracker behaviour for $P\ge 11$.
It also tends to produce lower $w_Q$ than the
inverse-power-law potential---for example
for $\Omega_Q\approx 0.7$, we have $w_Q(0) \approx -0.8$.
We also consider a quintessence model with a potential given
by the sum of two power laws,
\begin{equation}
V(Q) = \frac{A}{Q^{P_1}} + \frac{B}{Q^{P_2}}
\label{eqn:contrived}
\end{equation}
where $P_2 = P_1\times 10^{-6}$.  Because the powers of the two terms
differ by six orders of magnitude, this potential tends to produce
values for the equation of state that are very close to $-1$ at the
present epoch---when $\Omega_Q \approx 0.7$, we have $w_Q(0)\approx
-0.98$.  This type of potential was introduced by \cite{SWZ99} to show
that a very contrived potential is necessary for a tracker model to
produce values of $w_Q(0)$ very close to $-1$.  We use this potential
here to explore the power of SNAP to distinguish quintessence from
$\Lambda$ models in the extreme case when $w_Q(0)\rightarrow -1$.
Hereafter, we shall refer to simple inverse-power-law quintessence as
a PL model, to the supergravity-inspired quintessence model as a SUGRA
model, and to the sum-of-two-power-laws model as a 2PL model.
 
\begin{figure}
\begin{center}
\epsfig{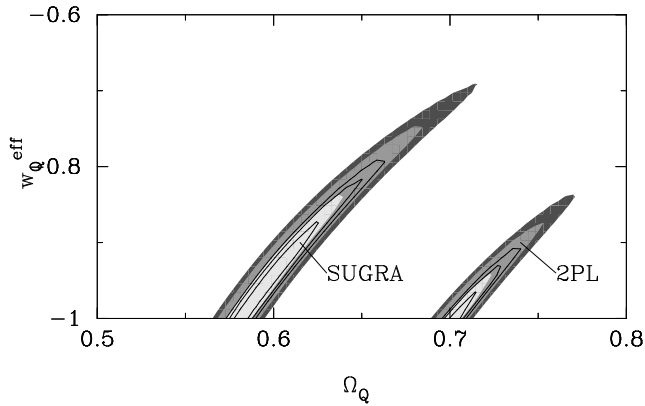}
\caption[Likelihood contours for $\Omega_Q$ and $w_Q^{\mathit{eff}}$,
for two different quintessence models]{Similar to 
Fig.~\ref{fig:2pcont}, except that
here the likelihood contours are for data drawn from tracker models
arising from the supergravity-inspired potential 
[``SUGRA''---equation~(\ref{eqn:SUGRA})] and the 
sum of two inverse power laws [``2PL''---equation~(\ref{eqn:contrived})].  
In neither case is it
possible, using SNAP data alone, to distinguish the quintessence
model in question from a $\Lambda$ model.}
\label{fig:2pcont_neg}
\end{center}
\end{figure}

Fig.~\ref{fig:2pcont_neg} shows the likelihood contours that
we obtain from SNAP-like datasets drawn from each of these
models.  In neither case is it possible to constrain 
$w_Q^{\mathit{eff}}\neq -1$; hence it is not possible to distinguish
either of these models from a cosmological constant using SNAP
data alone. (Our results for the $\Lambda$, PL, and SUGRA models
are in good agreement with the results of 
\cite{WA01,WA01-2}.)  
Apparently SNAP data, on their own, can only be used to rule 
out $\Lambda$ models in cases where $w_Q$ differs significantly
from $-1$.   In section~\ref{sec:cosfit}, though,
we found that applying a prior probability distribution on
$\Omega_Q$ helped to reduce the errors on all the parameters of our 
fitting function.
Using such prior information might help us to distinguish a larger
set of quintessence models from a $\Lambda$ model.  We shall explore this
possibility next.

But first we note that there is a 
discrepancy between our quoted values of $\Omega_Q$ and $w_Q(0)$
and the likelihood contours shown in figures~\ref{fig:2pcont}
and~\ref{fig:2pcont_neg}.  For example, for the SUGRA
model, we get present-day parameters $\Omega_Q \approx 0.7$ and
$w_Q(0) \approx 0.8$, but the likelihood analysis gives best-fitting
parameters of $\Omega_Q\approx 0.6$ and $w_Q^{\mathit{eff}}\approx -0.95$.
Similarly, for the PL model, the actual parameters
are $\Omega_Q \approx 0.7$ and $w_Q(0)\approx -0.6$, but the
best-fitting parameters are $\Omega_Q\approx 0.6$ and 
$w_Q^{\mathit{eff}}= -0.7$.  

It is important to keep in mind, though, that
equation~(\ref{eqn:rfit_2par}) is nothing more than a fitting function
to the coordinate distance. 
In this context, $\Omega_Q$ and $w_Q^{\mathit{eff}}$ are merely
fitting parameters, and our likelihood analysis simply finds the parameters
that produce the best fit to the data. 
Because our fitting function is only an approximation
to $\tilde{r}(z)$,   the best-fitting parameters may turn out to 
be significantly different from the actual physical parameters
they represent.  (In the present example, the best fit is given
by reducing both $w_Q(0)$ and $\Omega_Q$ from their actual values.)  
Nevertheless, these parameters could still be used to 
rule out a cosmological constant. In the case of a $\Lambda$ model, 
the fitting function
is an \emph{exact} representation of $\tilde{r}(z)$, so
we would expect the best-fit 
parameters to match the actual parameters.  Thus if we rule out $w_Q = -1$, 
we can say definitively that we are \emph{not} dealing with a 
$\Lambda$ model.  But it is important to recognize that we can say nothing 
definitive about the actual values of the  parameters, since the
best-fit parameters do not necessarily reproduce them.

\begin{figure}
\begin{center}
\epsfig{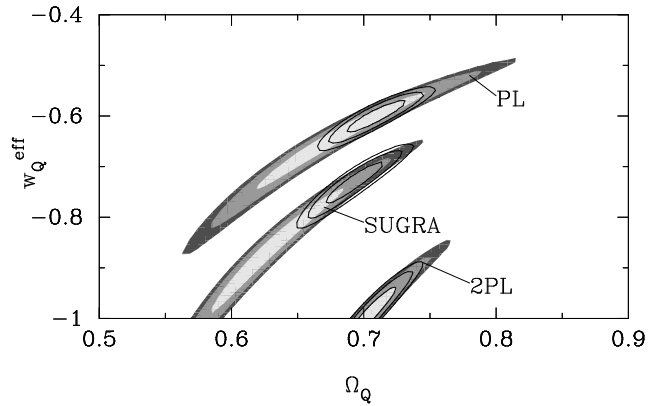}
\caption{Likelihood 
contours in $\Omega_Q$-$w_Q^{\mathit{eff}}$
space for various quintessence models, assuming Gaussian 
prior probability distributions on $\Omega_M = 1-\Omega_Q$.  The shaded 
contours indicate $68\%$, $95\%$ and $99\%$ confidence levels assuming a 
prior with standard deviation $\sigma_{\Omega_M} = 0.05$, and the line
contours define the same confidence regions for $\sigma_{\Omega_M} = 0.015$.
All contours use a one-year SNAP dataset.  Including such priors in
our analysis allows us to distinguish the PL and 
SUGRA models
from a cosmological constant, but they provide no discriminatory power
in the case of the 2PL model.}  
\label{fig:2pcont_sig}
\end{center}
\end{figure}

Now we return to the idea of imposing prior probability distributions
on our likelihood analysis.  As in section~\ref{sec:cosfit}, we
assume that we have measured $\Omega_M=1-\Omega_Q$ 
with Gaussian error $\sigma_{\Omega_M}$ about its true value.   
In view of the issues just discussed, it is important to 
say clearly what we are constraining here.  The maximum likelihood
parameters in this case are the parameters $\Omega_Q$ and $w_Q^{\mathit{eff}}$
that would be most likely to produce the SNAP data if $w_Q$ were
constant, given 
what we already know about $\Omega_Q$.  Because we would still expect these
most likely parameters to be the correct parameters in the case of a
$\Lambda$ model, ruling out $w_Q^{\mathit{eff}} = -1$  is
still tantamount to ruling out the cosmological constant.    
As shown Figure~\ref{fig:2pcont_sig}, then, it is possible to rule out a 
$\Lambda$ model with $99\%$ confidence for the case of a PL quintessence 
model, by combining a year of SNAP data and a measurement of $\Omega_M$ with 
$\sigma_{\Omega_M} = 0.05$.  In the case of the SUGRA
model, we can rule out the cosmological constant at $99\%$ confidence
if we have $\sigma_{\Omega_M} = 0.015$.  This is encouraging:  it appears
that using SNAP in concert with other cosmological observations will
provide a strong test of the basic properties of dark energy.

It is important to note, however, that the data cannot distinguish
the 2PL quintessence model
from a cosmological constant model, even with a very good measurement
of $\Omega_M$.   Although this model is highly contrived, it is clear
that SNAP will not be able to distinguish any arbitrary quintessence model
from a $\Lambda$ model using equation~(\ref{eqn:rfit_2par}) as a fitting
function.  As $w_Q$ approaches $-1$, quintessence becomes increasingly 
difficult
to distinguish from a cosmological constant model, and our only
hope for distinguishing quintessence from $\Lambda$ is to probe the 
evolution of $w_Q$ with redshift.

\subsection{Three-parameter fits}
\label{sec:like_3p}

\begin{figure}
\begin{center}
\epsfig{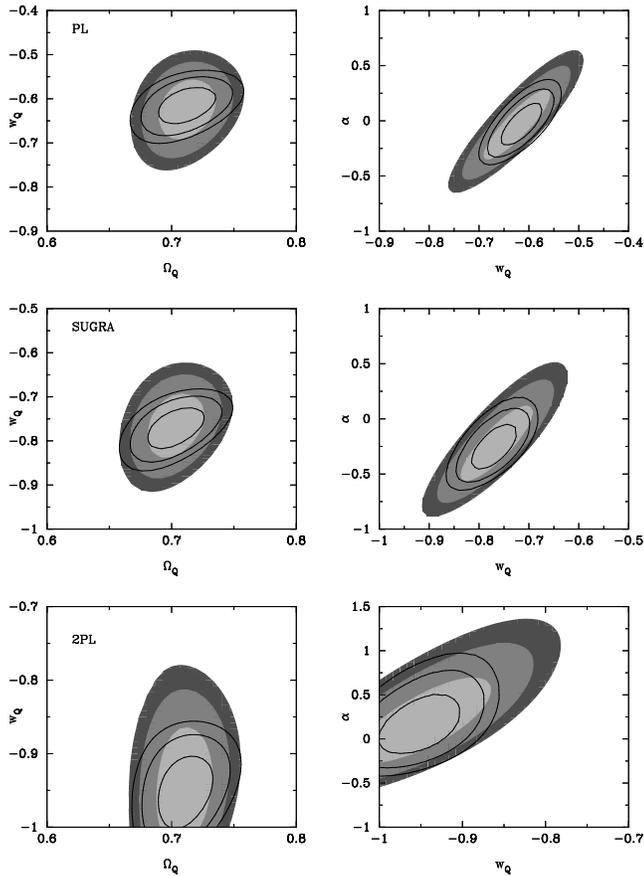}
\caption[Likelihood contours for $\Omega_Q$, $w_Q(0)$ and $\alpha$,
for various quintessence models with strong priors on $\Omega_M$]{Likelihood 
contours for the parameters $\Omega_Q$,
$w_Q$ and $\alpha$, as defined in 
equation~(\ref{eqn:fit_wln}).  The contours shown
are for, from top to bottom, the PL, SUGRA and 2PL models.
The shaded contours are for one year of SNAP data and the
line contours are for three years of data, where in each case
a Gaussian prior with $\sigma_{\Omega_M} = 0.015$ has been
imposed on $\Omega_M = 1-\Omega_Q$. 
Only in the case of the SUGRA model is it 
possible to constrain $\alpha$ to be nonzero at any significant
confidence level, and then only for a full, three-year dataset.}  
\label{fig:sigcont}
\end{center}
\end{figure}

To probe the evolution of $w_Q$ with redshift, we return to the 
three-parameter fitting function that we used previously for 
reconstruction, equation~(\ref{eqn:fit_wln}), which was based
on the approximate equation of state 
$w_Q^{app}(z) = w_Q(0) - \alpha\ln(1+z)$. As we saw in 
Fig.~\ref{fig:contour_P2},
to get a tight
constraint on $\alpha$  it was necessary to 
impose a Gaussian prior on $\Omega_M$ with an optimistic value for 
$\sigma_{\Omega_M}$ of something like $0.015$.  
Fig.~\ref{fig:sigcont} shows the likelihood contours
that would be expected for one and three years of SNAP data combined with
such a prior.   We consider the same three
quintessence models as before. Only in the case of the SUGRA model is it
possible to constrain $\alpha$ to be nonzero---and here only
at $68\%$ confidence, and only for a full, three-year dataset.

It is hardly surprising that $\alpha\neq 0$ is more readily shown
for the SUGRA model.  In this model, $w_Q$ is 
evolving relatively rapidly at at late times, with $\alpha=-0.43$,
as opposed to $-0.11$ for the PL model and 
$-0.01$ for the 2PL model.  Clearly, SNAP
will be much less powerful for distinguishing evolving from
non-evolving dark energy models than it is for distinguishing
quintessence from $\Lambda$ models.  In most cases, then, it appears
that  two-parameter fitting functions will be more useful than three-parameter
fitting functions for constraining cosmology with supernova data.

\section{Long-term prospects for probing quintessence}
\label{sec:ltparam}

So far we have seen that SNAP will be a powerful tool for determining
properties of the dark energy. But even with very good prior information,
it may not be possible to distinguish between two very similar-looking
models, such as a 2PL model and a $\Lambda$ model.
It has been shown  that supernova data do a
better job of constraining some cosmological parameters when the
survey is extended to higher redshifts, $z>2$ \citep{GPE99}. 
Although SNAP will only track
supernovae out to redshifts of $z=1.7$, it could be possible in the 
long term to supplement SNAP data with higher-redshift data from powerful
new telescopes like the Next Generation Space Telescope (NGST).  
To get some rough idea of the usefulness of such observations, 
we simulate them by adding a
sample of 30 supernovae at redshift $z=3$ to our SNAP-like datasets.

\begin{figure}
\begin{center}
\epsfig{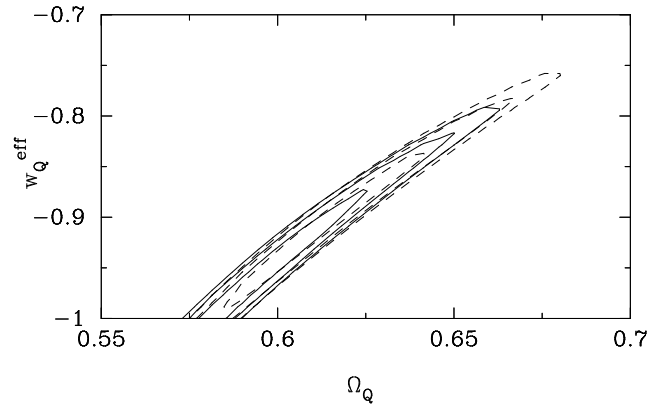}
\caption[Likelihood contours for $\Omega_Q$ and $w_Q^{\mathit{eff}}$,
with supplementary high-redshift observations]{Likelihood contours 
in $\Omega_Q$-$w_Q^{\mathit{eff}}$
space for the SUGRA quintessence model.
Solid contours define $68\%$, $95\%$ and $99\%$ confidence levels
for three years of SNAP data, and dashed lines define the same
regions when that data is supplemented by a sample of 30 supernovae
at redshift $z=3$.  When the high-redshift sample is included, 
it is possible to rule out a $\Lambda$ model ($w_Q^{\mathit{eff}}=-1$)
with $68\%$ confidence.  No prior knowledge of the parameters is 
assumed in either case.}
\label{fig:2pcont_lt}
\end{center}
\end{figure}

Fig.~\ref{fig:2pcont_lt} shows the improvement that can be made on
a two-parameter likelihood analysis for the SUGRA model, 
when we supplement SNAP data
with high-redshift observations.  Whereas previously
it was impossible to rule out a $\Lambda$ model using SNAP data 
alone, as shown in Fig.~\ref{fig:2pcont_lt}, 
when we supplement a three-year SNAP dataset with high-redshift
supernovae, we can rule out a $\Lambda$ model with $68\%$ confidence,
with no prior knowledge of the parameters. 
We see a similar shift in the confidence regions for the PL and 
2PL models, although the 2PL model 
still cannot be distinguished from a $\Lambda$ model.

\begin{figure}
\begin{center}
\epsfig{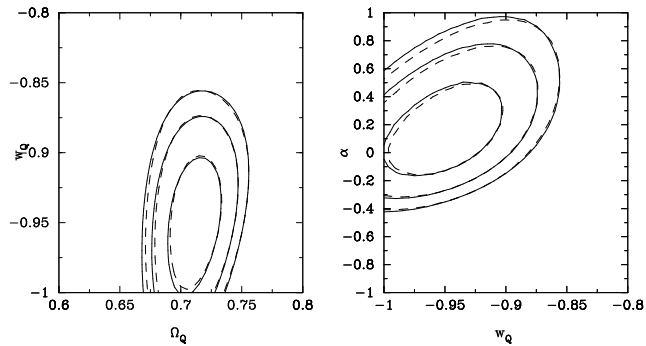}
\caption[Likelihood contours for $\Omega_Q$, $w_Q(0)$ and $\alpha$,
with supplementary high-redshift observations]{Likelihood contours for 
a three-parameter fit to SNAP-like
data drawn from a 2PL quintessence model.  Solid contours are for
three years of SNAP data; dashed contours are for a three-year dataset
supplemented with 30 supernovae at $z=3$.  With the high-redshift
sample, it is possible to distinguish the 2PL model from a $\Lambda$ model.}
\label{fig:ltcont_2PL}
\end{center}
\end{figure}

In the case of a three-parameter fit, the effect of adding a
high-redshift sample is smaller, but it can still be important
in some cases.  Fig.~\ref{fig:ltcont_2PL} shows the likelihood
contours that would be expected in the case of the 2PL quintessence
model, for a three-parameter fit to a three-year dataset, 
imposing a Gaussian prior on $\Omega_M$ with $\sigma_{\Omega_M} = 0.015$.   
The high-redshift observations do not affect the confidence contours 
strongly enough to indicate a nonzero $\alpha$, but they do constrain
$w_Q(0) \neq -1$ at $68\%$ confidence, ruling out a $\Lambda$ model
at that confidence level.  Obtaining this result is somewhat surprising,
since it was not possible to do so with a two-parameter fit.
The reason it is possible here is that, since our three-parameter fitting 
function 
allows $w_Q$ to vary with $z$, the parameters $w_Q(0)$ and $\Omega_Q$
are likely to be more accurately constrained, whereas
the two-parameter fit required an offset in the best-fit parameters
to make up for the $z$-dependence of $w_Q$. Three-parameter fitting 
functions, then, may be more useful than two-parameter
fitting functions in some cases, even when they cannot be used to 
discern the evolution of $w_Q$.

Finally, we note that Fig.~\ref{fig:ltcont_2PL} assumes a very
optimistic observational scenario.  To distinguish the 
2PL model from a $\Lambda$ model, we must have three years of
SNAP data with negligible systematic errors, plus thirty high-redshift
supernovae from NGST, an independent measurement of $\Omega_M$
with $\sim 5\%$ errors, and extremely good evidence that the universe 
is flat.  But although an observational programme that would lead to 
constraints like those in Fig.~\ref{fig:ltcont_2PL} is ambitious, it
is certainly not impossible.  The simulated SNAP dataset we used in this
analysis has realistic statistical errors, and systematics are expected
to be negligible in comparison \citep{SNAP}.  NGST, when launched, will
almost certainly see at least some additional high-redshift supernovae.
Also, as we have noted, our assumed prior constraints on $\Omega_M$ are
realistic.  Furthermore, the 
MAP and Planck satellites should provide very precise measurements
of the curvature of the universe [to within a percent or so for Planck
\citep{Planck}].  The first
steps have been taken toward observations that could lead to a measurement
like Fig.~\ref{fig:ltcont_2PL}.  Although this figure represents
a best-case scenario, it also represents an achievable
long-term goal.
 
\section{Conclusion}
\label{sec:conclusion}

In this study we have investigated the prospects for probing the dark 
energy, particularly quintessence, with observations of type Ia supernovae.  
We have examined tests for individual quintessence models (reconstruction)
as well as tests that address more general questions about the 
dark energy (parameter estimation by likelihood analysis).  In both cases, 
we find that data from the proposed SNAP satellite will
provide important information about dark energy, either alone
or when combined with other cosmological observations. 

In the case of reconstruction it will be important to take 
extreme care in choosing a function to fit SNAP's measurement
of the distance-redshift relation $r(z)$. 
By using a fitting function based on a physical approximation
to the coordinate distance, it is possible to produce accurate and 
reliable reconstructions from SNAP data combined with independent 
cosmological observations. Such reconstructions might provide useful
observational constraints as we construct models for the dark energy.

To answer our basic questions about the nature of the dark
energy, however, cosmological parameter estimation may be more useful.  
If the dark energy equation of state is 
significantly different from $-1$, the SNAP data alone will be able
to rule out the cosmological constant.  Moreover, combining the SNAP
constraints with independent measurements of $\Omega_M$ will allow
us to rule out $\Lambda$ models for all but the most extreme cases of 
quintessence, when $w_Q$ is very close to $-1$.  
SNIa are less likely to provide evidence that $w_Q$ is evolving. 
Even combining SNAP data with very precise measurements of
$\Omega_M$ will only provide evidence of an evolving $w_Q$ in cases
of extremely strong evolution.   Three-parameter fits may still be useful, 
however:  because they allow a more accurate determination of $w_Q(0)$
than two-parameter fits, they may be able to rule out a cosmological
constant in cases where a two-parameter fit cannot do so.

The prospects are  good 
for probing dark energy with SNIa data.
SNAP, by itself, may be able to constrain
cosmological parameters with enough precision to rule out the
cosmological constant as a dark energy candidate. However, 
it is when SNIa data are combined with other observations that their true 
value becomes apparent.  When combined with expected future measurements
of $\Omega_M$, for example, SNAP data provide much stronger constraints
on the quintessence equation of state and may be able to 
produce useful reconstructions of the equation of
state and quintessence potential.  
In each of these cases, the combination of several observations
tells us much more than any of the observations by itself, and
in each case, the SNIa observations are an essential component.  The
measurements that may answer our questions about dark energy 
over the next decade or two constitute a difficult and impressive
observational programme. Improved supernova observations are 
crucial to its success.

\section*{Acknowledgments}
BFG acknowledges the generous support of the Herchel Smith fellowship 
from Williams College, USA. 
%
%

%
%
\bibliography{quint.mn}

\begin{thebibliography}{}

\bibitem[\protect\citeauthoryear{Armendariz, Mukhanov \& Steinhardt}{Armendariz
  et~al.}{2000}]{AMS00}
Armendariz C.,  Mukhanov V.,    Steinhardt P.,  2000, PRL, 85, 4438

\bibitem[\protect\citeauthoryear{Armendariz, Mukhanov \& Steinhardt}{Armendariz
  et~al.}{2001}]{AMS01}
Armendariz C.,  Mukhanov V.,    Steinhardt P.,  2001, PRD, 63, 103510

\bibitem[\protect\citeauthoryear{Astier}{Astier}{2001}]{Astier00}
Astier P.,  2001, Phys. Lett. B, 500, 8

\bibitem[\protect\citeauthoryear{Baccigalupi et~al.,}{Baccigalupi
  et~al.}{2002}]{Baccigalupi01}
Baccigalupi et~al., 2002, PRD, 65, 063520

\bibitem[\protect\citeauthoryear{Barger \& Marfatia}{Barger \&
  Marfatia}{2001}]{BM00}
Barger V.,  Marfatia D.,  2001, Phys. Lett. B, 498, 67

\bibitem[\protect\citeauthoryear{Bean}{Bean}{2001}]{Bean01}
Bean R.,  2001, PRD, 64, 123516

\bibitem[\protect\citeauthoryear{Bean \& Melchiorri}{Bean \&
  Melchiorri}{2002}]{BM02}
Bean R.,  Melchiorri A.,  2002, PRD, 65, 041302

\bibitem[\protect\citeauthoryear{Brax \& Martin}{Brax \& Martin}{1999}]{SUGRA}
Brax P.,  Martin J.,  1999, Phys. Lett. B, 468, 40

\bibitem[\protect\citeauthoryear{Caldwell, Dave \& Steinhardt}{Caldwell
  et~al.}{1998}]{CDS98}
Caldwell R.,  Dave R.,    Steinhardt P.~J.,  1998, PRL, 80, 1582

\bibitem[\protect\citeauthoryear{Chiba \& Nakamura}{Chiba \&
  Nakamura}{2000}]{CN00}
Chiba T.,  Nakamura T.,  2000, PRD, 62, 121301

\bibitem[\protect\citeauthoryear{Corasanti \& Copeland}{Corasanti \&
  Copeland}{2002}]{CC01}
Corasanti P.,  Copeland E.,  2002, PRD, 65, 043004

\bibitem[\protect\citeauthoryear{de Bernardis et~al.,}{de~Bernardis
  et~al.}{2002}]{Bernardis01}
de Bernardis P.,  et~al., 2002, ApJ, 564, 559

\bibitem[\protect\citeauthoryear{Efstathiou}{Efstathiou}{1999}]{GPE99}
Efstathiou G.,  1999, MNRAS, 310, 842

\bibitem[\protect\citeauthoryear{Efstathiou et~al.,}{Efstathiou
  et~al.}{1999}]{GPEetal99}
Efstathiou G.,  et~al., 1999, MNRAS, 303, L47

\bibitem[\protect\citeauthoryear{Efstathiou et~al.,}{Efstathiou
  et~al.}{2002}]{GPEetal02}
Efstathiou G.,  et~al., 2002, MNRAS, 330, L29

\bibitem[\protect\citeauthoryear{Eisenstein, Hu \& Tegmark}{Eisenstein
  et~al.}{1998}]{EHT98}
Eisenstein D.,  Hu W.,    Tegmark M.,  1998, ApJ, 504, L57

\bibitem[\protect\citeauthoryear{Goliath et~al.,}{Goliath
  et~al.}{2001}]{Goliath01}
Goliath M.,  et~al., 2001, A\&A, 380, 6

\bibitem[\protect\citeauthoryear{Haiman, Mohr \& Holder}{Haiman
  et~al.}{2001}]{HMH01}
Haiman Z.,  Mohr J.,    Holder G.,  2001, ApJ, 553, 545

\bibitem[\protect\citeauthoryear{Hu et~al.,}{Hu  et~al.}{1998}]{Huetal98}
Hu W.,  et~al., 1998, PRD, 59, 023512

\bibitem[\protect\citeauthoryear{Huterer \& Turner}{Huterer \&
  Turner}{1999}]{HT99}
Huterer D.,  Turner M.,  1999, PRD, 60, 081301

\bibitem[\protect\citeauthoryear{Huterer \& Turner}{Huterer \&
  Turner}{2001}]{HT00}
Huterer D.,  Turner M.,  2001, PRD, 64, 123527

\bibitem[\protect\citeauthoryear{Levi et~al.,}{Levi  et~al.}{2000}]{SNAP}
Levi M.,  et~al., 2000, http://snap.lbl.gov

\bibitem[\protect\citeauthoryear{Maor, Brustein \& Steinhardt}{Maor
  et~al.}{2001}]{MBS01}
Maor I.,  Brustein R.,    Steinhardt P.,  2001, PRL, 86, 6

\bibitem[\protect\citeauthoryear{Nakamura \& Chiba}{Nakamura \&
  Chiba}{1999}]{NC99}
Nakamura T.,  Chiba T.,  1999, MNRAS, 306, 696

\bibitem[\protect\citeauthoryear{Netterfield et~al.,}{Netterfield
  et~al.}{2002}]{boom}
Netterfield C.,  et~al., 2002, ApJ

\bibitem[\protect\citeauthoryear{Peebles \& Ratra}{Peebles \&
  Ratra}{1988}]{PR88}
Peebles P. J.~E.,  Ratra B.,  1988, ApJ, 325, L17

\bibitem[\protect\citeauthoryear{Perlmutter et~al.,}{Perlmutter
  et~al.}{1999}]{SCP}
Perlmutter S.,  et~al., 1999, ApJ, 517, 565

\bibitem[\protect\citeauthoryear{Perlmutter, Turner \& White}{Perlmutter
  et~al.}{1999}]{PTW99}
Perlmutter S.,  Turner M.,    White M.,  1999, PRL, 83, 670

\bibitem[\protect\citeauthoryear{Pryke et~al.,}{Pryke  et~al.}{2002}]{dasi}
Pryke C.,  et~al., 2002, astro-ph/0104490

\bibitem[\protect\citeauthoryear{Riess et~al.,}{Riess  et~al.}{1998}]{Hi-z}
Riess A.,  et~al., 1998, AJ, 116, 1009

\bibitem[\protect\citeauthoryear{Saini et~al.,}{Saini  et~al.}{2000}]{Saini00}
Saini T.,  et~al., 2000, PRL, 85, 1162

\bibitem[\protect\citeauthoryear{Starobinsky}{Starobinsky}{1998}]{Starobinsky9%
8}
Starobinsky A.,  1998, JETP Lett., 68, 757

\bibitem[\protect\citeauthoryear{Steinhardt, Wang \& Zlatev}{Steinhardt
  et~al.}{1999}]{SWZ99}
Steinhardt P.,  Wang L.,    Zlatev I.,  1999, PRD, 59, 123504

\bibitem[\protect\citeauthoryear{Tauber et~al.,}{Tauber  et~al.}{2000}]{Planck}
Tauber J.,  et~al., 2000, http://astro.estec.esa.nl/SA-general/Projects/Planck

\bibitem[\protect\citeauthoryear{van Waerbecke, Bernardeau \& Mellier}{van
  Waerbecke et~al.}{1999}]{WBM99}
van Waerbecke L.,  Bernardeau F.,    Mellier Y.,  1999, A\&A, 342, 15

\bibitem[\protect\citeauthoryear{Vilenkin}{Vilenkin}{2001}]{Vilenkin01}
Vilenkin A.,  2001, hep-th/0106083

\bibitem[\protect\citeauthoryear{Wang, Tegmark \& Zaldarriaga}{Wang
  et~al.}{2002}]{WTZ01}
Wang X.,  Tegmark M.,    Zaldarriaga M.,  2002, PRD

\bibitem[\protect\citeauthoryear{Wang \& Garnavich}{Wang \&
  Garnavich}{2001}]{WG01}
Wang Y.,  Garnavich P.~M.,  2001, ApJ, 552, 445

\bibitem[\protect\citeauthoryear{Wang \& Lovelace}{Wang \&
  Lovelace}{2001}]{WL01}
Wang Y.,  Lovelace G.,  2001, ApJ, 562, L115

\bibitem[\protect\citeauthoryear{Weller \& Albrecht}{Weller \&
  Albrecht}{2001}]{WA01}
Weller J.,  Albrecht A.,  2001, PRD, 86, 1939

\bibitem[\protect\citeauthoryear{Weller \& Albrecht}{Weller \&
  Albrecht}{2002}]{WA01-2}
Weller J.,  Albrecht A.,  2002, PRD

\bibitem[\protect\citeauthoryear{Weller, Battye \& Kneissl}{Weller
  et~al.}{2001}]{WBK01}
Weller J.,  Battye R.,    Kneissl R.,  2001, astro-ph/0110353

\bibitem[\protect\citeauthoryear{Zlatev, Wang \& Steinhardt}{Zlatev
  et~al.}{1999}]{ZWS99}
Zlatev I.,  Wang L.,    Steinhardt P.,  1999, PRL, 82, 896

\end{thebibliography}
\bibliographystyle{mn2e}
\end{document}